# Radio interference reduction in interstellar communications: methods and observations

*William J. Crilly Jr.*

Green Bank Observatory, West Virginia, USA

***Abstract*— The discovery of interstellar communication signals is complicated by the presence of radio interference. Consequently, interstellar communication signals are hypothesized to have properties that favor discovery in high levels of local planetary radio interference. A hypothesized type of interstellar signal, Δ*t* Δ*f* polarized pulse pairs, has properties that are similar to infrequent elements of random noise, while dissimilar from many types of known radio interference. Discovery of Δ*t* Δ*f* polarized pulse pairs is aided by the use of interference-filtered receiver systems that are designed to indicate anomalous presence of polarized pulse pairs, when pointing a radio telescope to celestial coordinates of a hypothetical transmitter. Observations reported in previous work (ref. arXiv:2105.03727) indicate an anomalous celestial pointing direction having coordinates 5.25 ± 0.15 hours Right Ascension and -7.6° ± 1° Declination. Augmented interference reduction mechanisms used in the current work are described, together with reports of follow-up radio telescope beam transit measurements during 40 days. Conclusions and further work are proposed.**

## I. Introduction

The experimental protocol in previous work [1] produced measurements of the Additive White Gaussian Noise (AWGN)-caused likelihood of observations of energy-efficient [2] interstellar coherence hole (ICH) constrained signals, described by Messerschmitt [3]. The hypothesis in [1] predicts that indications at celestial coordinates 5.25 ± 0.15 hours Right Ascension (*RA*) and -7.6° ± 1° Declination (*DEC*), may be explained by an AWGN model. The experiment described in [1] appears to have falsified the hypothesis to the extent measured by the Bayesian posterior probabilities calculated in [1]. The absence of development and testing of auxiliary and alternate hypotheses hampers conclusions. The problem compels the work reported in this paper.

Follow-on work, proposed in [1], to continue single telescope beam transit observations, attempts to further study certain hypothesized ICH-constrained signals, referred to as Δ*t* Δ*f* discovery signals, and polarized pulse pairs. During the course of the follow-on work, four additional Radio Frequency Interference (RFI) amelioration mechanisms were designed and implemented, as additional suspected RFI was observed and excised, and while hypothetical interstellar communication methods were considered.

William J. (Skip) Crilly Jr. is a Volunteer Science Ambassador in Education & Public Outreach of the Green Bank Observatory. email: wcrilly@nrao.edu

Due to the change in experimental protocol in the follow-on work, the hypothesis stated in [1] is retained, albeit modified for the current beam transit experiment.

***Hypothesis* [1]:** An AWGN-cause model is expected to explain observations of Δ*t* Δ*f* orthogonal circular-polarized pulse pairs, narrow bandwidth ICH-constrained elements of hypothetical interstellar transmitted signals, while a radio telescope is pointed on and off celestial coordinates 5.25 ± 0.15 hours Right Ascension (*RA*) and -7.6° ± 1° Declination (*DEC*). The term *polarized pulse pair* is used in this work to refer to the hypothetical interstellar signals.

In one argument, the Bayesian posterior probabilities calculated in the current work may include the factor of prior probabilities, posteriors calculated in [1]. In another argument, prior probabilities may be considered as indications of the significance of the 5.25 ± 0.15 hours *RA* and -7.6° ± 1° *DEC* parameters placed in the predictive hypothesis above. The hypothesis may then be subjected to a test with new experimental data, with no explicit Bayesian prior, or with the inclusion of the prior, depending on assumptions.

The remainder of this paper describes four methods of RFI reduction, added to the methods described in [1], summarizes 40-day beam transit observations using the 26 foot radio telescope in New Hampshire, and discusses results, conclusions, and proposed further work.

## II. Methods

The RFI filter mechanisms that excise candidate polarized pulse pairs, based on Δ*t*, Δ*f* and polarized pulse pair interarrival time, are described in **Figure 1**.

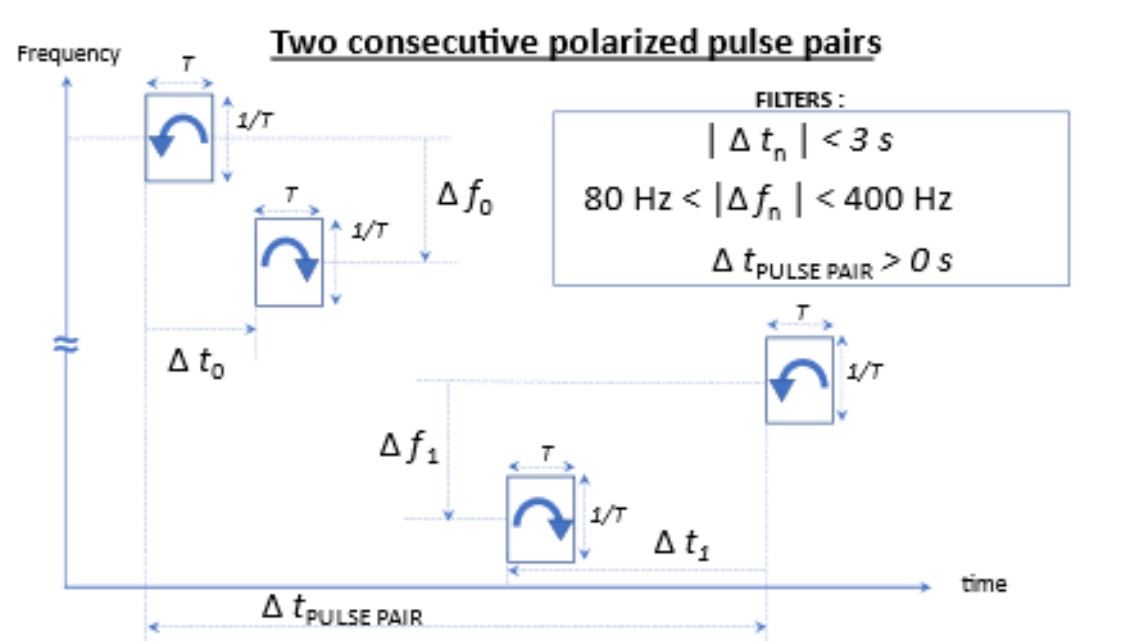

**Figure 1**: Filters are designed to reduce RFI-caused indications of hypothetical interstellar polarized pulse pair signals. 1/*T* is equal to 3.725 Hz in this work.

In the augmented RFI filters, measured $\Delta t$ and $\Delta f$ are constrained to limits, and the interarrival time of candidate polarized pulse pairs, $\Delta t_{\text{PULSE PAIR}}$, is required to be non-zero. RFI amelioration methods implemented in machine post-processing [1], and retained in the current work, are summarized below, together with the description of the four added or changed amelioration methods implemented in the current work.

1. Telescope-specific, pre-observation, spectral segment excision of persistent RFI,
2. Post-processing spectral excision of persistent RFI,
3. Dynamic RFI excision using spectral Infinite Impulse Response filtering of Signal-to-Noise Ratio (SNR) threshold crossings,
4. Harmonically related frequency excision, within ±25 kHz of 500 kHz harmonics, and ±1 kHz of 100 kHz harmonics, (the latter implemented, but not stated in [1]),
5. Observation frequency band edge and in-phase quadrature (IQ) near-zero baseband excision.
6. **Changed:** $\Delta t$ $\Delta f$ polarized pulse pairs measuring $|\Delta t|$ less than 3 seconds are included as candidate polarized pulse pairs. Previous work [1] used $\Delta t = 0$ throughout the machine post-processing of observations. Two reasons explain the use of $\Delta t = 0$ in [1]. Synchronized geographically-spaced radio telescope observations in [1] naturally lead to the search for $\Delta t = 0$ polarized pulse pairs, implying a single pulse is propagated to both telescopes. Further, $\Delta t = 0$ polarized pulse pairs have minimum likelihood in AWGN, yet have potentially high flux due to RFI. The increase of the $|\Delta t\,|$ filter maximum to 3 seconds is hypothesized to increase the non-RFI polarized pulse pair flux, improving statistical properties during relatively short-term beam transit measurements.
7. **Changed:** $\Delta t$ $\Delta f$ polarized pulse pairs having $|\Delta f\,| < 80$ Hz, are excised, to reduce indications caused by Doppler-spread RFI. Doppler spread of signals is caused by various movements of transmitting antennas, receiving antennas, and various movements of objects located within primary Fresnel zones of the signal propagation paths [4].
8. **Added:** Consecutive $\Delta t$ $\Delta f$ polarized pulse pairs having an inter-arrival time of zero are excised. Elements of RFI from a single transmitting source tend to be coincident, within the pulse integration period $T$, after propagation to the receiver. Possible leakage of RFI through the polarized pulse pair filters, i.e. RFI amelioration methods 1 through 7, may confuse experimental results. The use of a filter having $\Delta t_{\text{PULSE PAIR}} > 0$ is hypothesized to reduce RFI-caused indications, while reducing received true-positive polarized pulse pair flux by a minimal amount, as polarized pulse pairs are hypothesized to be infrequent when transmitted at a single pulse duration $T$, among other possible values of transmitted pulse duration.
9. **Changed:** Sorting of SNR, high to low, by the lower SNR of the two orthogonal circular polarizations' SNRs, is implemented in the current work. The SNRs of candidate polarized pulse pair elements have a threshold of 13.0 dB for the higher of the two polarized SNRs, and 11.8 dB for the lower SNR. In previous work [1], the higher SNR of the two orthogonal polarizations' SNRs was used in the sort, for reasons that follow. In receiver systems that filter RFI with $\Delta t = 0$, e.g. using geographically-spaced radio telescopes as in [1], a single high SNR pulse received at one telescope is compared to indications at the second telescope. Reduction of threshold SNR at the second telescope seems natural, when seeking a weak signal originating from a single transmitting source, propagated into each telescope's beam pattern. When using a single telescope in beam transit measurements, repetitive polarized RFI near $\Delta t = 0$ is expected to confuse measurement results. RFI is hypothesized to have polarization measurements localized on the Poincare´ sphere, with significant SNR difference indications of orthogonal polarizations. The flux of transmitted polarized pulse pairs is hypothesized to increase at high SNR thresholds of each orthogonal polarization, due to properties of the Ricean to Rayleigh density ratio, described in Appendix B of [1]. Sorting by the lower of the polarized SNRs therefore seems important in this work.

The RFI amelioration methods 2 – 9 above are implemented in the post-processing of telescope raw data. Telescope measurements of 11.8 dB SNR threshold crossing events are saved in files during signal capture, to reduce the risk of corruption of telescope experimental raw data due to post-processing methods.

*RA* filtering is implemented in beam transit measurements, similar to *RA* filtering used in [1], to correlate anomalous polarized pulse pair flux to a hypothetical celestial direction. In the current work, *RA* is quantized to 0.3-hour intervals, providing 80 *RA* intervals over 24 hours of *RA*, at -7.6° *DEC*. Binomial density likelihood functions use event probability equal to the ratio 0.3 hours *RA* / 24 hours *RA*, as uniform celestially distributed events are expected in AWGN.

## III. Observations

A 40-day beam transit measurement was conducted, beginning shortly after a 44-day artificial noise test was completed, the latter described in [1]. Observations were undertaken using the experimental protocol described in [1], modified by the augmented RFI filters, described in **II. Methods**. The post-processing and presentation of results follows procedures described in Appendix C, Method B of [1], with filter parameters, SNR sorting, and *RA* coverage modified as described in **II. Methods.**

**Figures 2 – 9** describe the binomial density likelihoods of SNR-sorted polarized pulse pairs, given an AWGN cause. Several anomalies are observed within the population of 80 *RA* ranges, suggesting that unexcised RFI might be present in the experimental results. Text below each figure summarizes anomalous measurements in the ten ranges of *RA* plotted in each figure.

In **Figure 3**, the 5.1 – 5.4 hour *RA* window indicates a significantly low binomial density calculated value, 0.00063, at 417 SNR trials, perhaps as a result of undiscovered RFI, equipment, and/or another cause. AWGN-caused pulse pair count likelihood, normalized using the expected AWGN-caused binomial density at 417 trials, calculates to 0.0036.

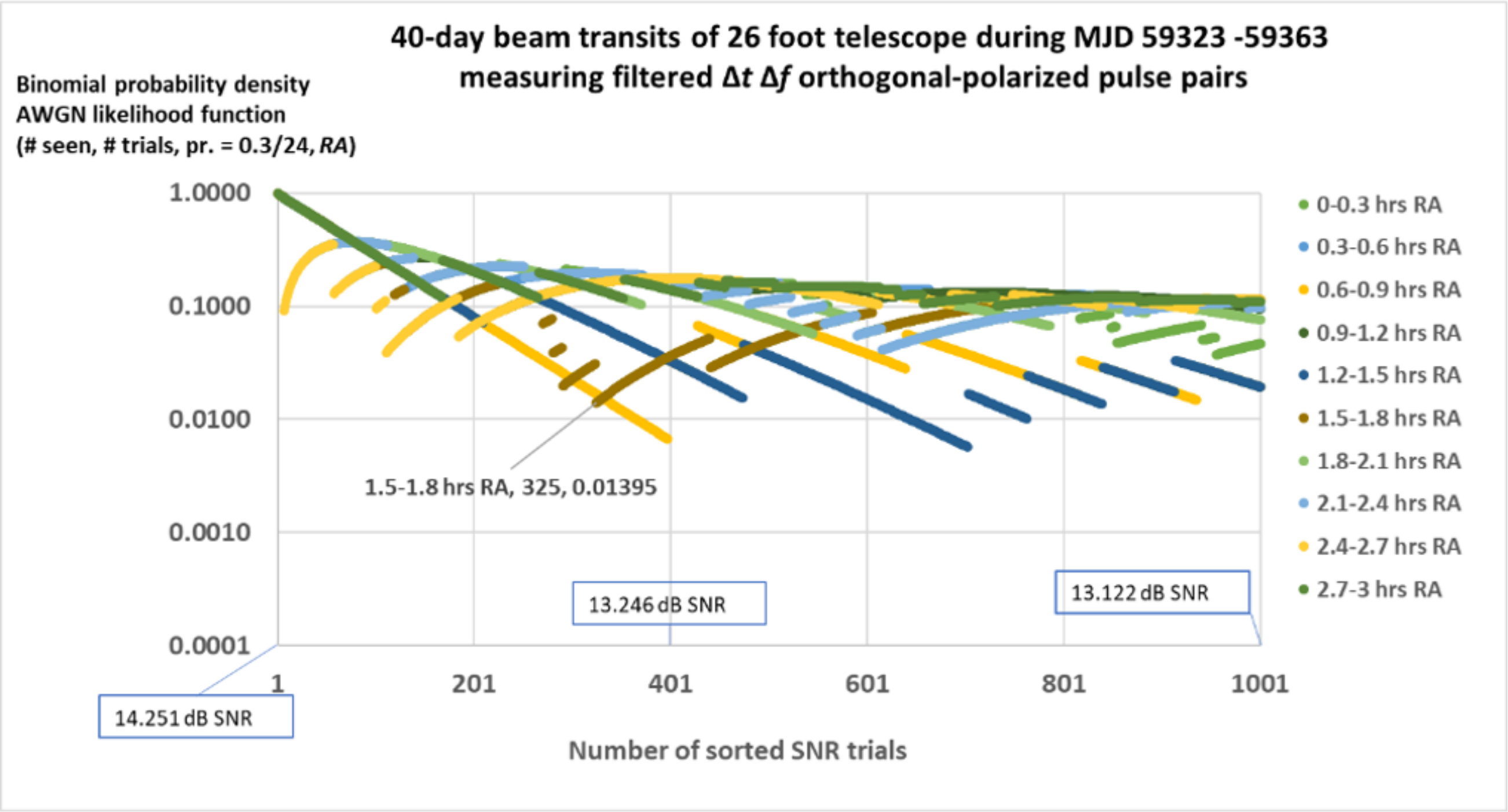

**Figure 2:** The 1.5 – 1.8 hour *RA* range indicates reduced binomial densities. Decreasing density discontinuities in **Figures 2 – 9** indicate anomalous presence of polarized pulse pairs, due to the two-sided binomial density function.

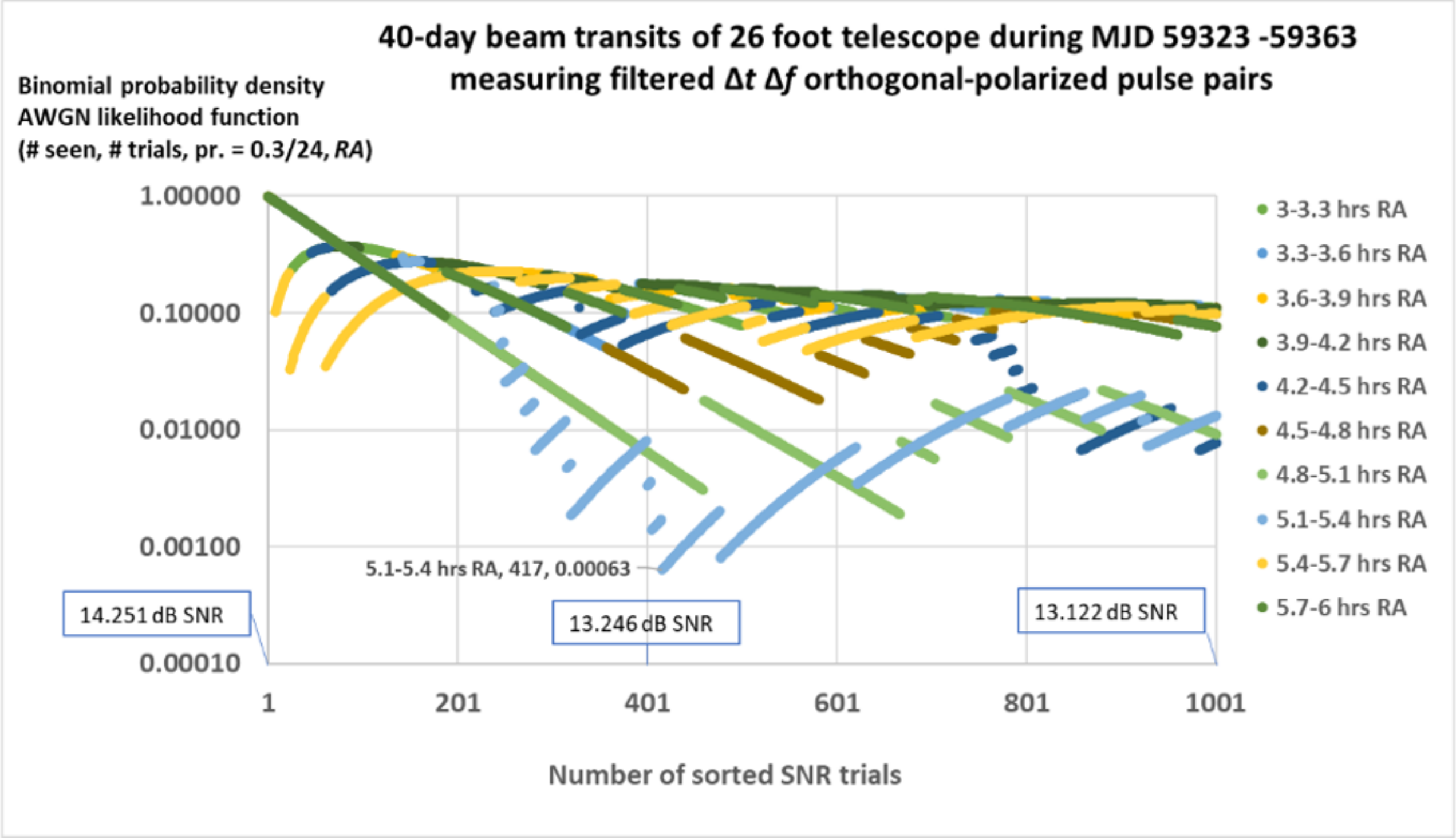

**Figure 3:** Polarized pulse pairs indicate in the 5.1 – 5.4 hour *RA* range, values of parameters of the hypothesis in this work, at a minimum of 0.00063 binomial density due to AWGN, 0.0036 times the density expected in an AWGN-only model at 417 trials. Fourteen polarized pulse pairs were observed, while a mean of 5.2 pulse pairs is expected in AWGN at the 417 SNR trial level. Pulse pairs appear distributed across Modified Julian Days (MJDs) and RF frequency, described in **Figure 12**. $\Delta f$ vs. $\Delta t$ are plotted in **Figure 13**. The 5.1 – 5.4 hour *RA* pointing direction indicates the third lowest binomial density, in 80 *RA* intervals. The lowest binomial density was observed in a group of anomalous pulse pairs primarily seen on MJD 59332, described in **Figures 9** and **10**.

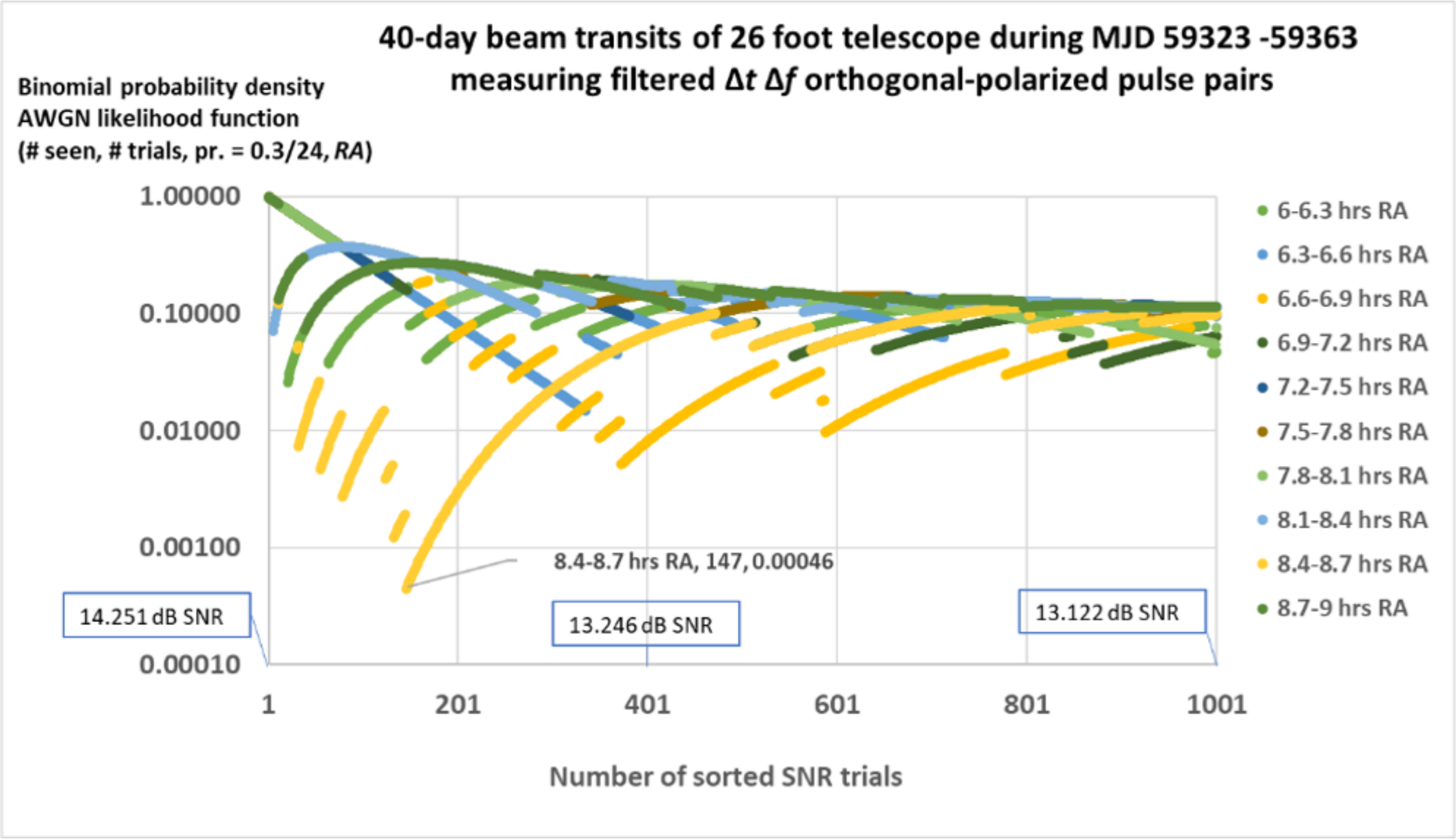

**Figure 4:** The 8.4 – 8.7 hour *RA* range indicates reduced binomial density of excess polarized pulse pairs. Distribution of polarized pulse pairs across MJD and RF Frequency are described in **Figure 11**.

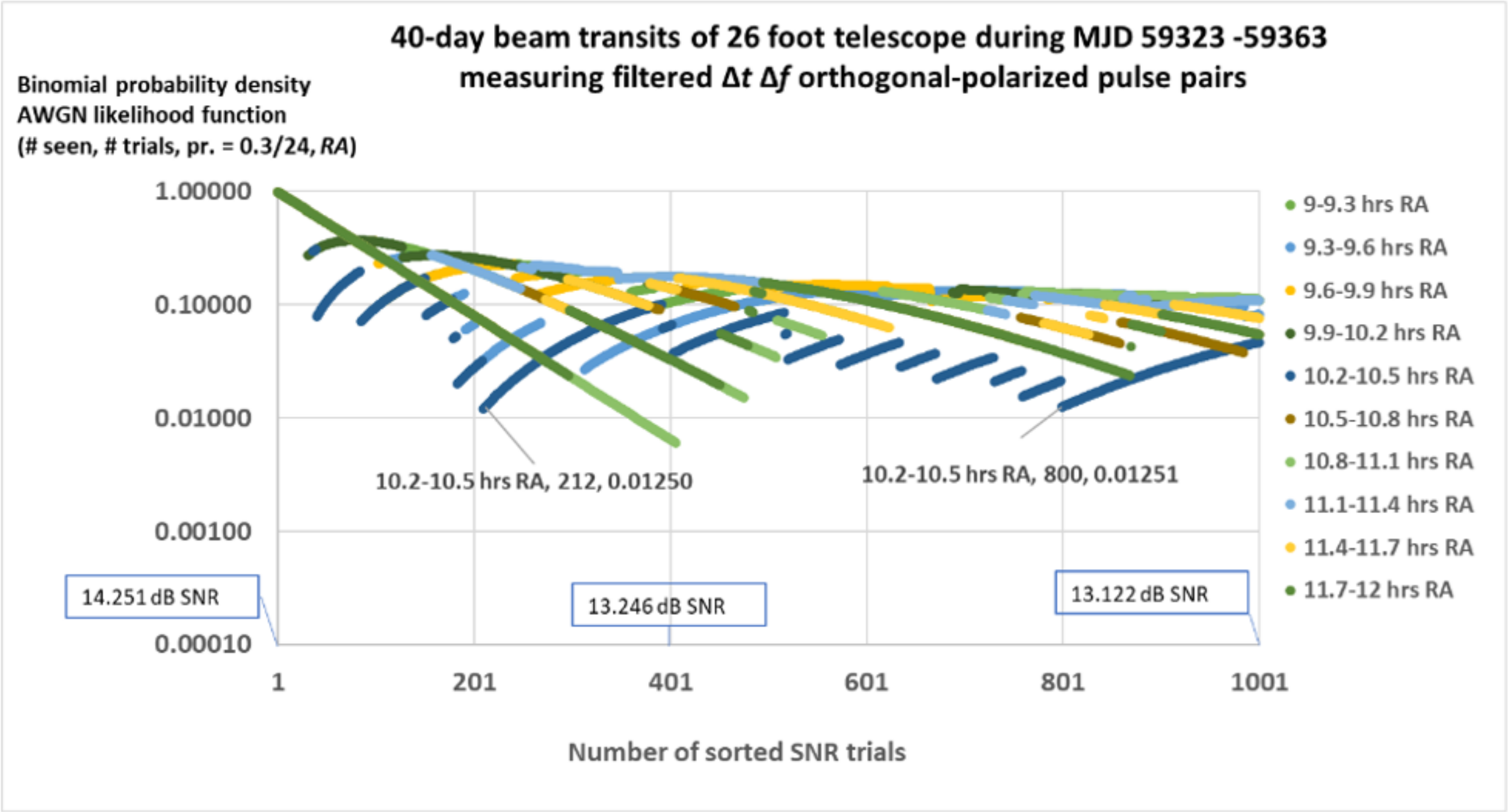

**Figure 5:** The 10.2 – 10.5 hour *RA* range indicates two SNR-sorted regions of reduced binomial density of excess polarized pulse pairs.

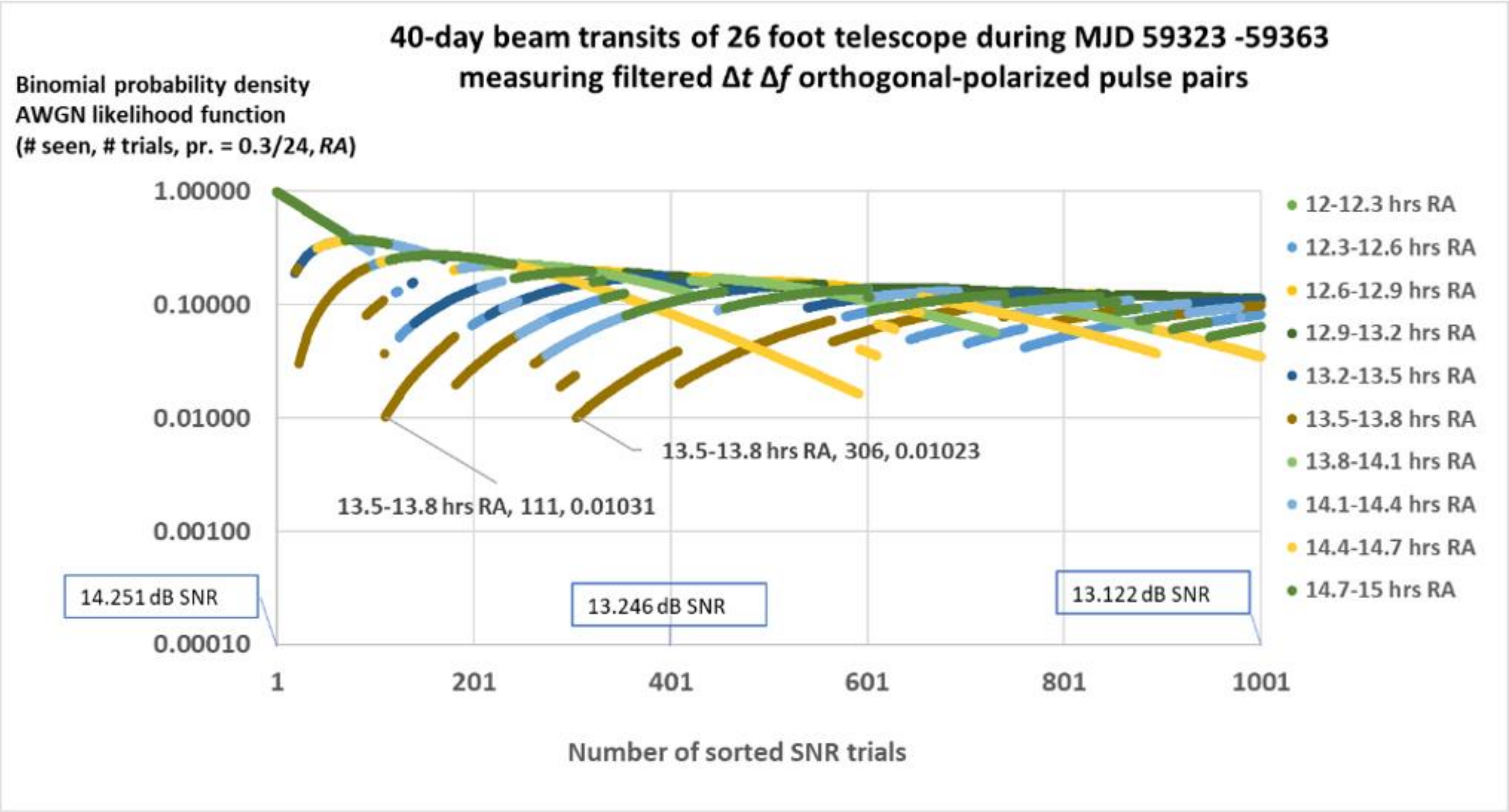

**Figure 6:** The 13.5 – 13.8 hour *RA* range indicates reduced binomial densities, displaying two indications at ≈ 0.01 density.

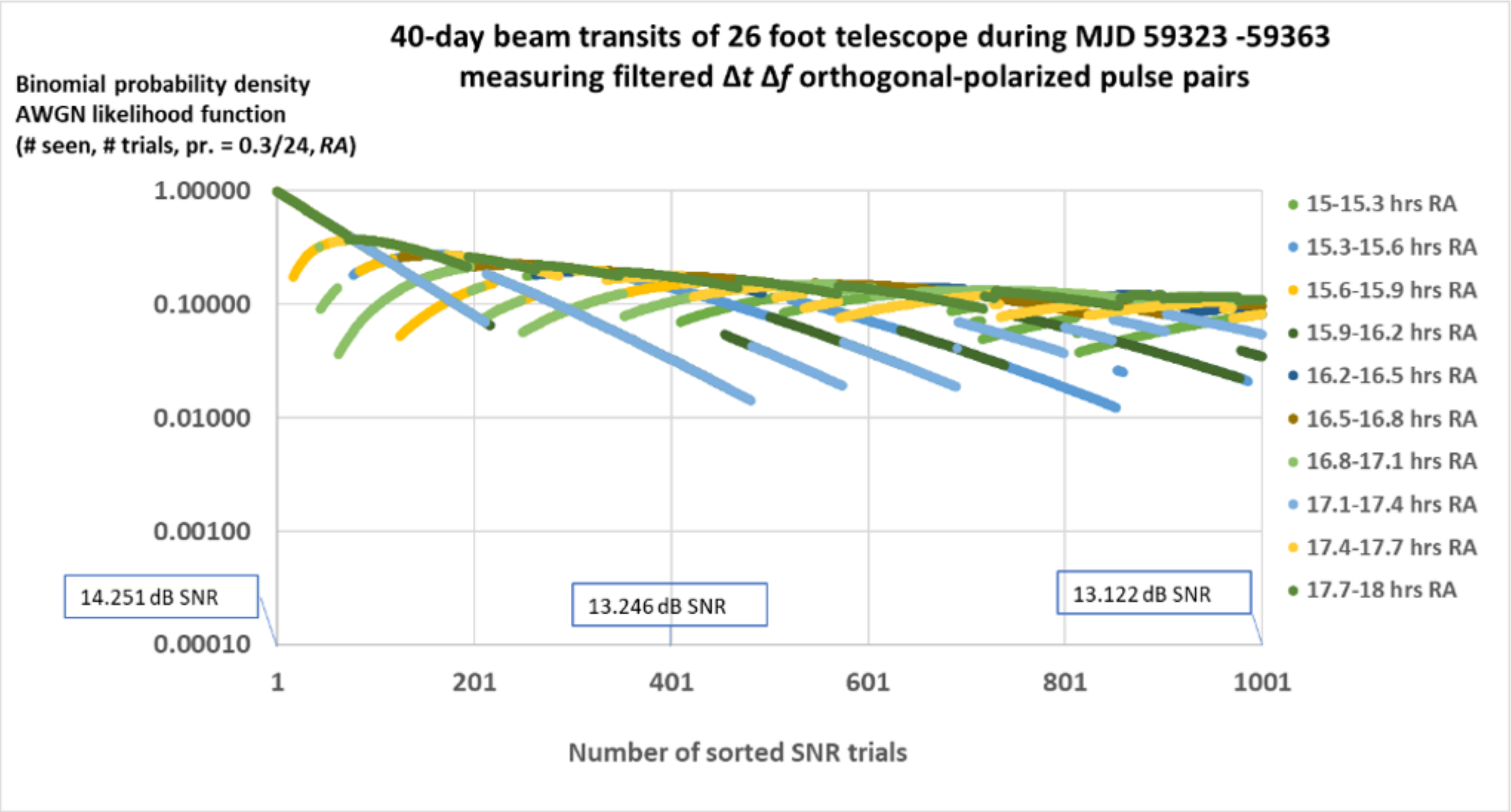

**Figure 7:** The 15 – 18 hour *RA* ranges do not indicate anomalous polarized pulse pairs.

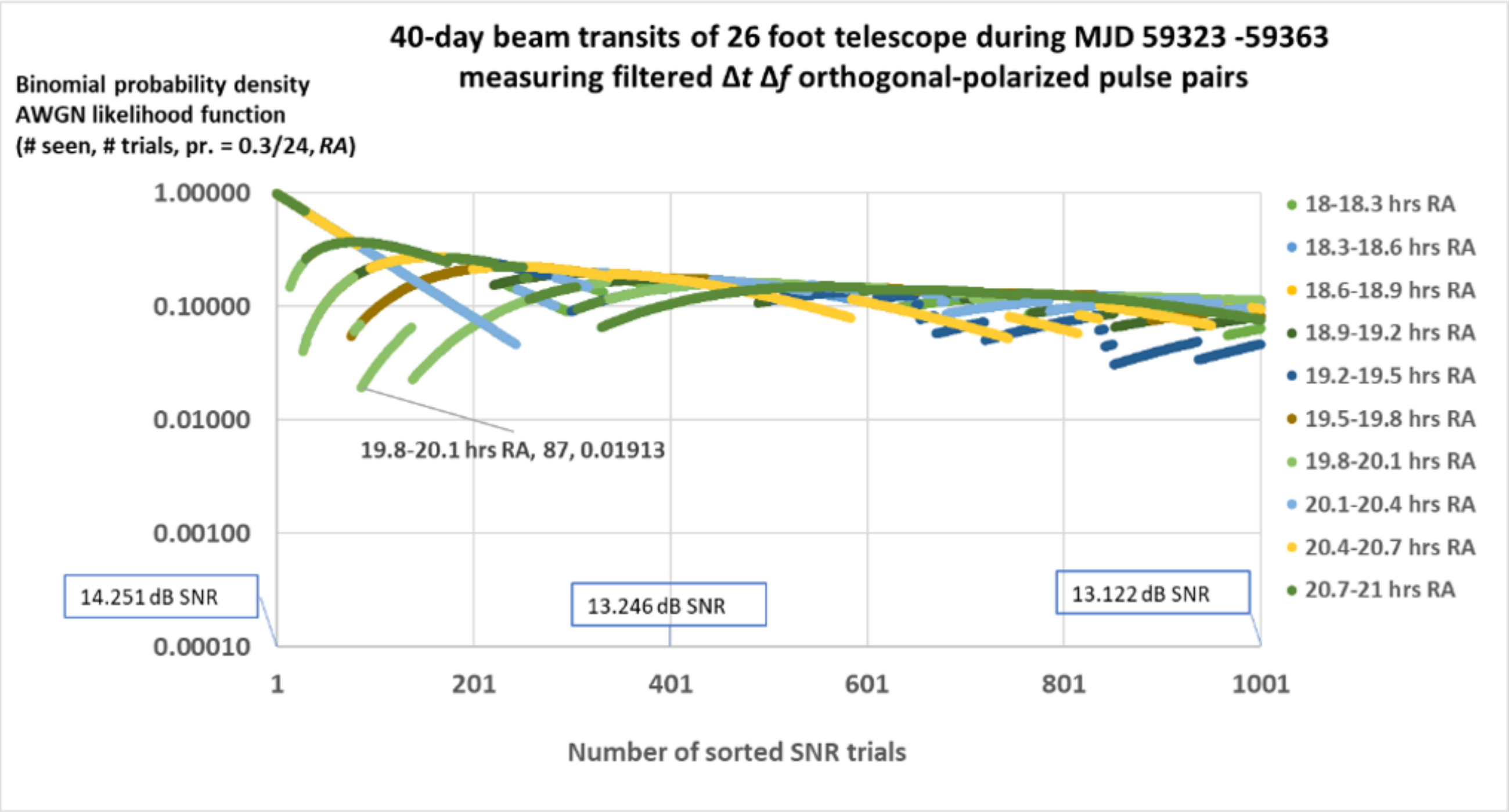

**Figure 8:** The 19.8 – 20.1 hour *RA* range indicates several high SNR binomial density reductions.

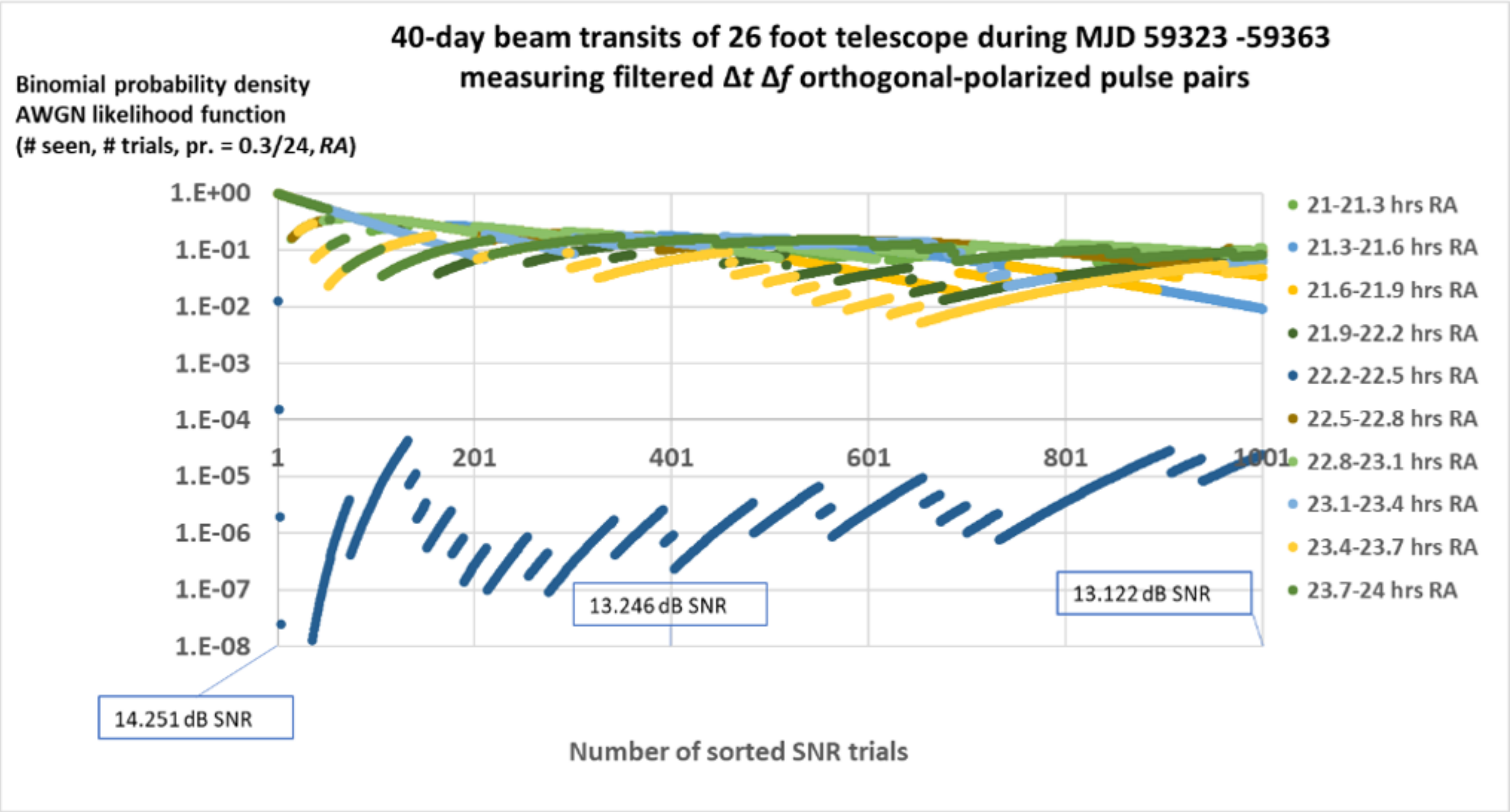

**Figure 9:** The 22.2 – 22.5 hour *RA* range anomalies significantly indicate on a single MJD day, MJD 59332, described in **Figure 10**. The single day presence of these pulse pairs, i.e. during a single beam transit, suggests a possible RFI explanation. Measured wideband distributed RF frequency of 22.2 – 22.5 hour *RA* polarized pulse pairs, similar to hypothetical interstellar signals, points to a potential undiscovered and unfiltered RFI source, possibly affecting data in other *RA* directions. A beam transit experiment is now underway (June 2021) to study alternate explanations.

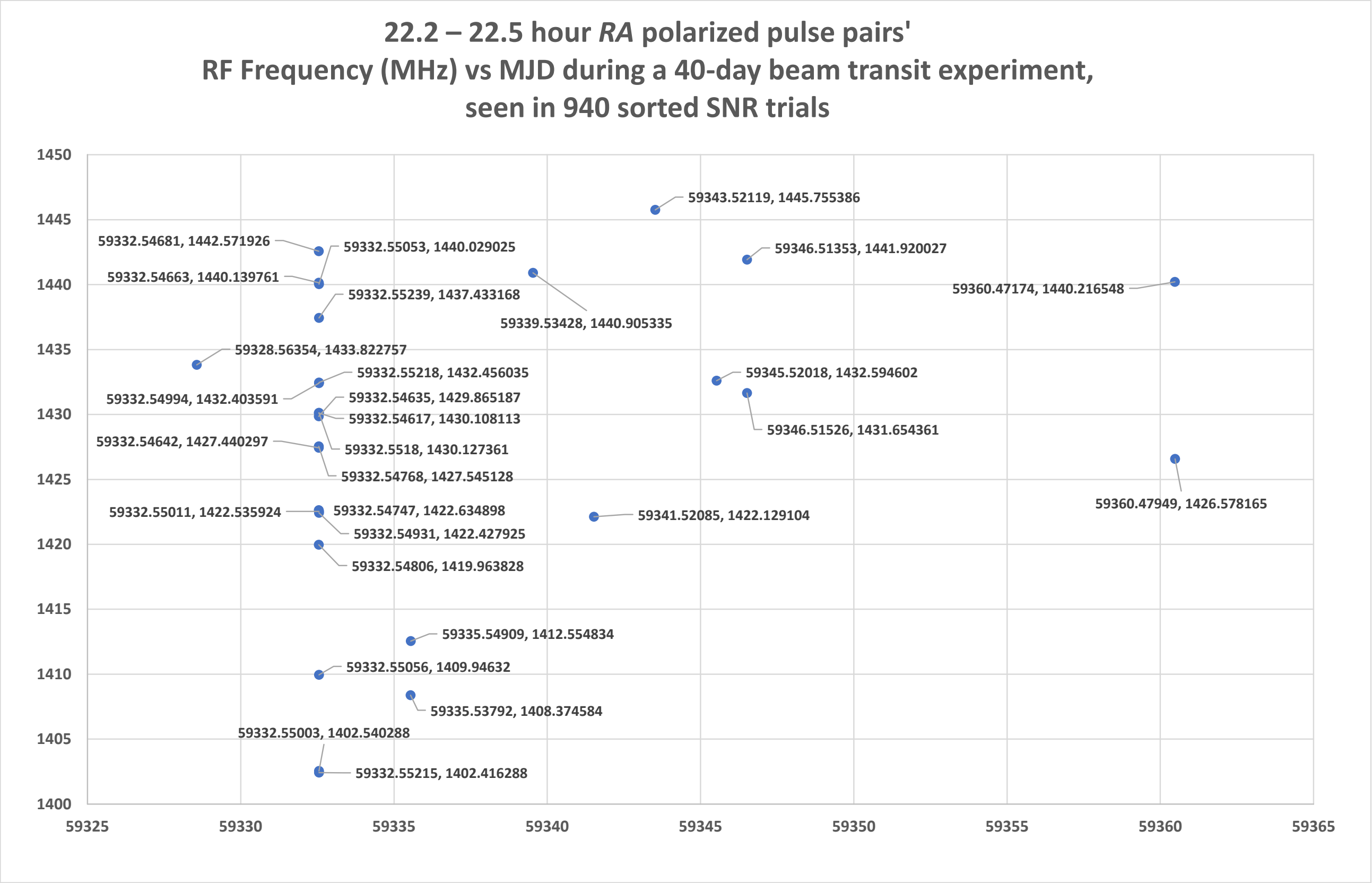

**Figure 10:** The 22.2 to 22.5 hours *RA* polarized pulse pairs' RF Frequencies, having measured MJD 59332, are concentrated with frequency differences near the fundamental and harmonics of approximately 2.44 MHz.

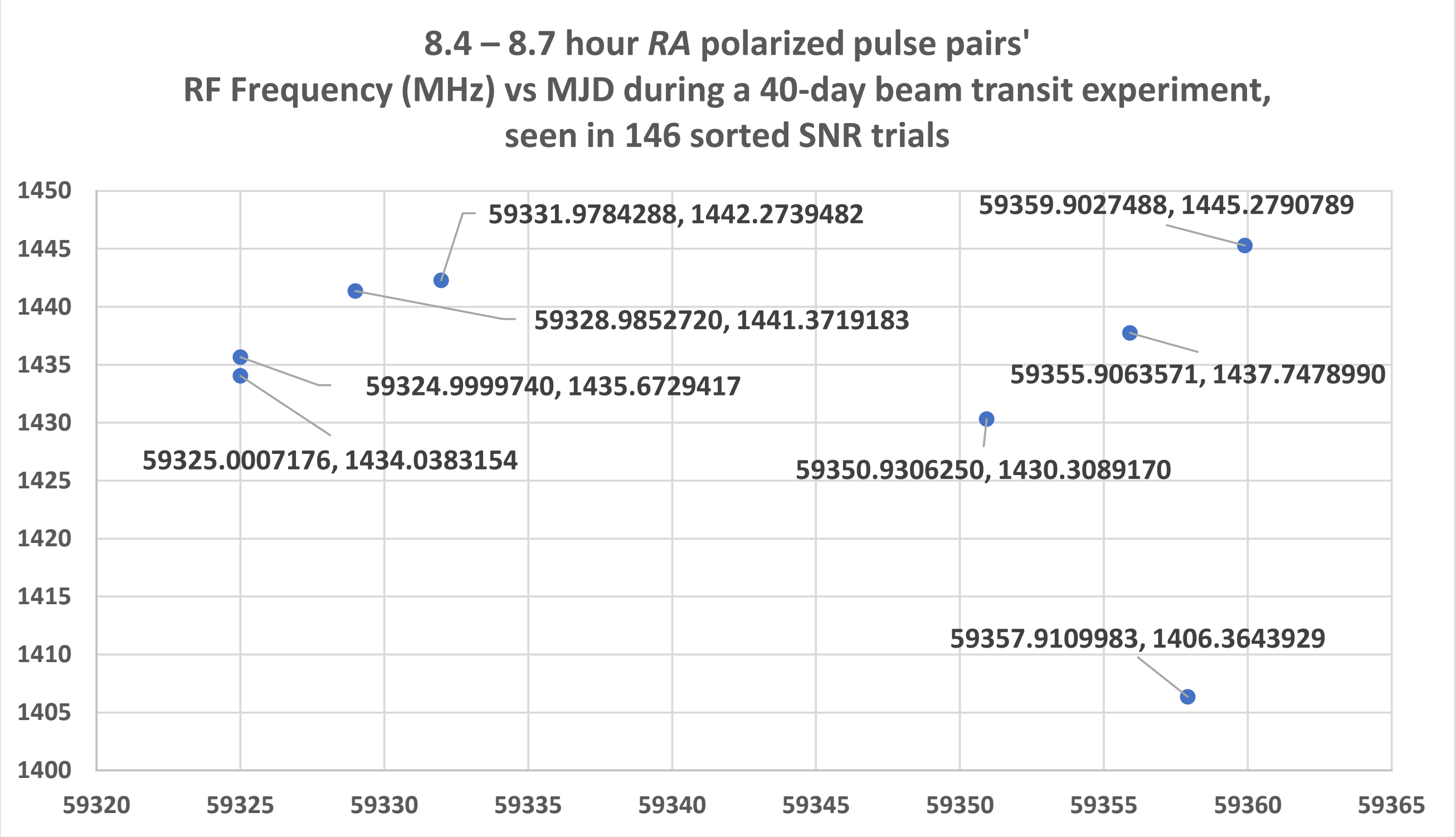

**Figure 11:** The 8.4 – 8.7 hour *RA* polarized pulse pairs appear distributed in MJD and RF Frequency.

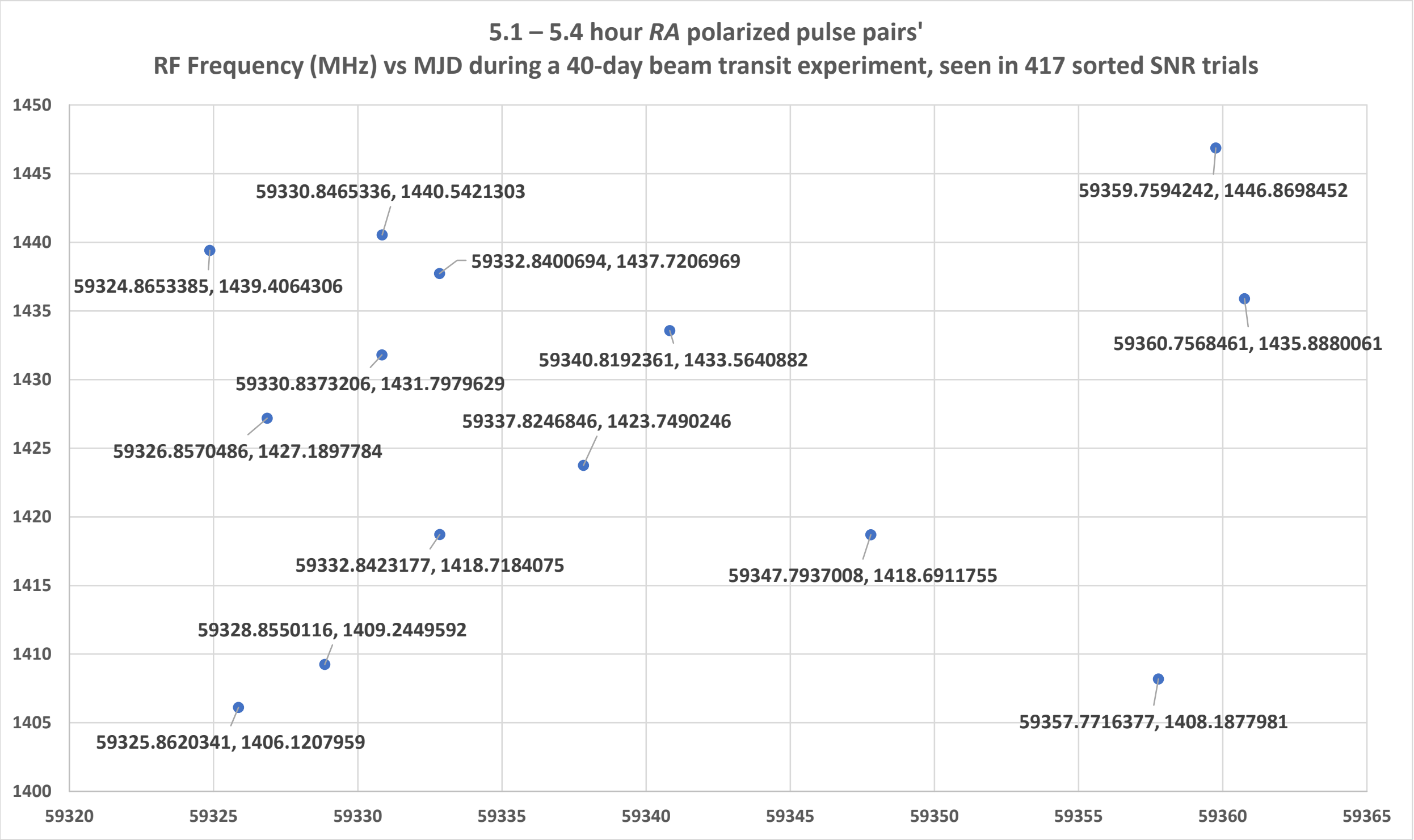

**Figure 12:** With the exception of the two polarized pulse pair events near 1418.7 MHz, anomalous 5.1 – 5.4 hour *RA* polarized pulse pairs appear distributed in MJD and RF frequency, as expected in an energy-efficient communications signal. Fourteen polarized pulse pairs were observed, while 5.2 polarized pulse pairs are expected, given an AWGN model.

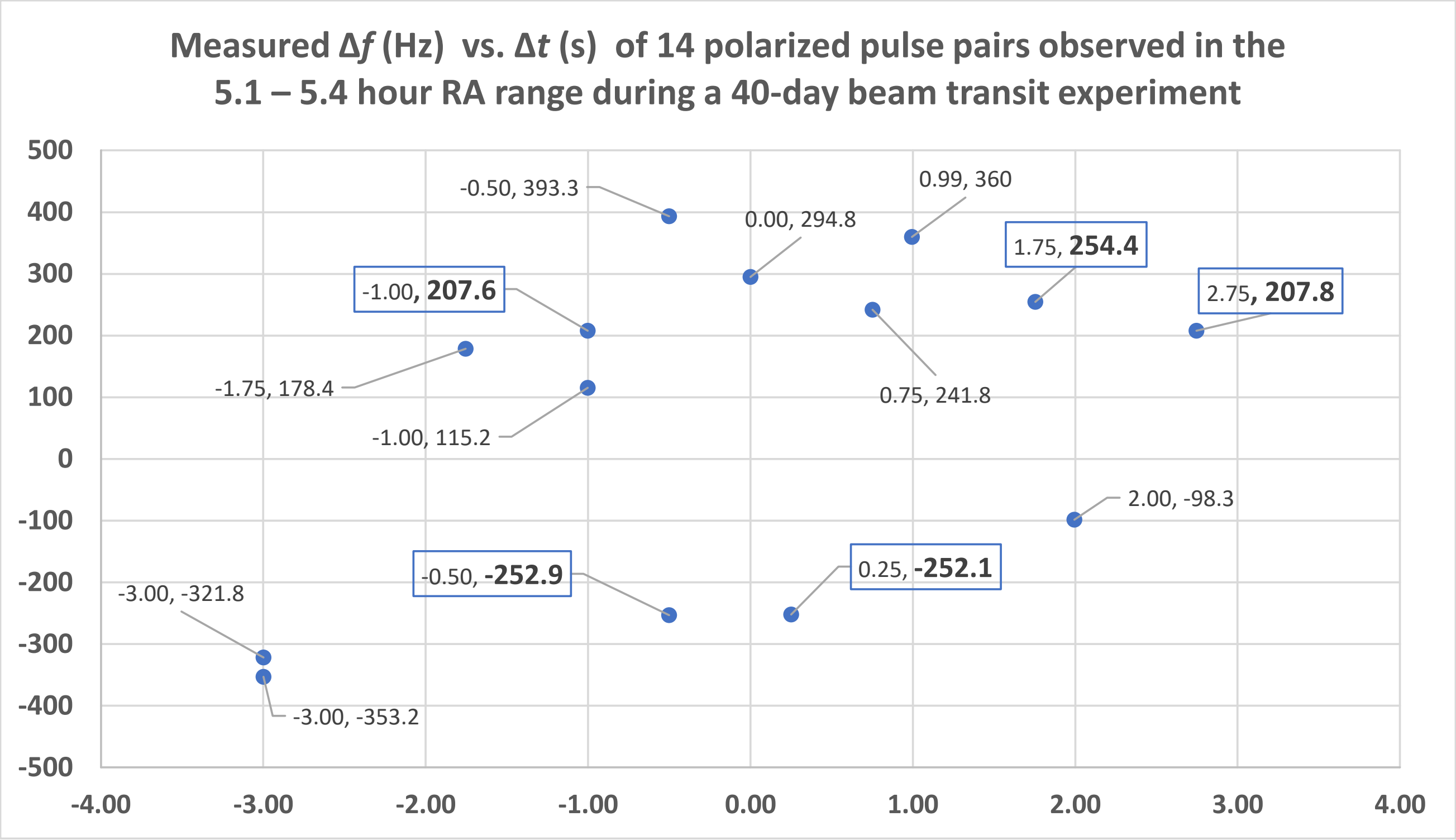

**Figure 13:** $\Delta f$ (Hz) **vs.** $\Delta t$ **(s)** of the 14 highest sorted SNR polarized pulse pairs in the 5.1 to 5.4 hour *RA* range. Four of the polarized pulse pairs indicate at nearly the same $\Delta f$ values, near 207.7 and -252.5 Hz. The probability density of similar matching $\Delta f$ pulse pair events, per experiment, calculates and simulates to 0.075, given AWGN. The probability density per experiment of the $\Delta f$ = 254.4 Hz polarized pulse pair, calculates to 0.19, given AWGN, assuming uniform $\Delta f$, and that a near-match $|\Delta f|$ = 254.4 Hz might measure within one FFT bin (3.7 Hz) of ±207.7 Hz, or ±252.5 Hz.

## IV. Discussion

In Bayesian inference, a tested model inherently includes aspects of the methods used to select the data population that is applied to the model's likelihood function [5].

The AWGN model stated in the hypothesis in this and previous work [1] is considered a model that has intentionally robust RFI filtering, in an attempt to select data that has the properties of AWGN, and consequently, hypothetical AWGN-like polarized pulse pairs. The resulting selection of a studied data population that is a subset of the overall telescope data may confuse the calculation of likelihoods.

To counter this confusion, an assumption is made that the theory underlying human-made RFI phenomena is captured in the algorithms and machine design that reject telescope data. In other words, RFI filters are considered experimental design aspects, similar to the reduction of telescope sidelobes, radio quiet zones, Faraday cages, etc. The model to be tested in this experiment is therefore an AWGN model that has intentionally rejected, in various ways, theoretical human-made radio interference.

The resulting model will be referred to as an AWGN model, given the argument that filters and equipment in an experiment are commonly designed to reject data that is understood to be not relevant, and ignorable to the discovery of the phenomena sought, i.e. AWGN-like communication signals. Design documentation and post-processing output files may be examined to determine if the experimental design and data population selection process are introducing false positives.

The binomial distribution is used in a likelihood function that estimates the probability of an AWGN cause of the measured count of polarized pulse pairs, within one of 80 *RA* ranges, at an SNR-sorted trial number. The underlying assumption that justifies the binomial is that each *RA* range is assumed to be independent of the 79 other *RA* ranges, allowing a binomial distribution to be applied separately to each *RA* range, instead of a multinomial distribution, comprising all ranges. The binomial assumption seems valid if the events in the other 79 *RAs* are adequately independent of events in the *RA* range examined, and adjacency and quantization of *RA* are ignorable.

The presence in the post-processing output file of apparent non-AWGN components, in one or more alternate *RA* ranges, within the binomial trials of a hypothesized significant *RA* range, biases the calculated binomial density likelihoods of the hypothetically significant *RA* range, with bias analyzed as follows. If one considers an intentional and justified elimination of $N$ suspected alternate *RA*, non-AWGN components, then the number of trials in the binomial calculation of a hypothetically significant *RA* range is reduced by $N$, reducing calculated likelihood, when pulse pair count is anomalously high. To prevent this type of bias, suspected alternate *RA* non-AWGN components are retained during binomial calculations, understanding that anomalous events in alternate *RA*s might bias otherwise low calculated likelihoods to slightly higher values. The use of the binomial likelihood seems justified if the number of alternate *RA* non-AWGN pulse pairs is low relative to the number of trials in the binomial calculation. This condition is often met when the number of *RA* ranges is large, and when many *RA* ranges appear noise-like.

The AWGN model, associated with its hypothesis and experimental method, appears to have been falsified in previous work, to a Bayesian posterior of approximately $10^{-4}$ (Figure 12 text in [1]). The current work might use this value as a prior, and calculate a Bayesian posterior at less than $10^{-6}$, as a result of the AWGN likelihood of 0.0036, calculated from data in **Figure 3,** using Bayesian inference methods described in Appendix E of [1]. Data invalidity arguably might increase this value significantly. For example, a fortuitous set of pulses may have been present in the 5.1 – 5.4 hour *RA* range, caused by noise, RFI, and/or equipment issues. If these complicating issues are assumed to be present, then the probability that data is valid may be low. The Bayesian posterior probabilities of many models, given invalid data, subsequently increase. Further, a combination of auxiliary and alternate hypotheses might explain the anomalies in **Figures 3, 4** and/or **9**.

A modified experimental step might invalidate the use of Bayesian inference, because, for example, a second model is considered in the Bayesian conditional likelihood function development, while a first model was used in the development of the prior probability. However, if an assumption is made that augmented RFI filters likely remove, and do not add, non-AWGN anomalies, then the use of a pre-augmentation model's prior in Bayesian inference might be justified, during a falsification. In another argument, almost any change in experimental protocol complicates a prior's subsequent use in Bayesian inference, due to uncertainty of the effect of the protocol change.

Robust RFI filtering presents the problem that hypothetical interstellar communication signals may be rejected by the RFI filters. An interesting outcome of the design and implementation of the robust RFI amelioration in this work is the inherent ability of the RFI filters to reject many types of suspected human-made RFI. Anomalously high counts of $\Delta t$ $\Delta f$ polarized pulse pairs might then appear, as either filter leakage of human-made RFI, or signals caused by something other than human-made RFI. In effect, the AWGN model, together with robust RFI filtering in the experimental protocol, seems useful to the discovery of intentionally transmitted $\Delta t$ $\Delta f$ polarized pulse pair signals.

In addition to equipment, methodology, and RFI leakage issues, one or more natural objects may explain anomalous results. In general, $T$ duration natural signals having relatively high levels of energy confined to bandwidth $1/T$ = 3.7 Hz are considered rare, due to the Doppler spread of rotating transmitting objects. Continuum measurements recorded during the 40-day beam transit test do not indicate a broadband telescope average power response in the range of 5.1 – 5.4 hours *RA*, at or above the sensitivity of the 26 foot telescope. Natural object measurement work is preliminary and does not include measurements and analysis that might indicate response due to various types of natural objects.

## V. CONCLUSIONS

The results of this experiment appear to falsify the AWGN model hypothesis, to the Bayesian posterior levels inferred from **III. OBSERVATIONS,** and **IV. DISCUSSION**. The AWGN model is falsified to a lesser extent if the prior from [1] is not considered. Models derived from auxiliary and alternate hypotheses have not been designed and tested to a degree necessary to propose conclusions that might explain the discrepancy between AWGN-caused expectations and observed results. Absent experimental methods, theory and evidence to test various alternate and auxiliary hypotheses, further work is important.

## VI. FURTHER WORK

The further work described in [1] is retained. Prioritization is adjusted to focus on the development of RFI models and tests. Spectrum analyzers with dedicated antennas have been added to the receiver system, continuously measuring suspected potential RFI sources. Beam transit measurements run almost continuously. Examination of the various reduced values of binomial density, RFI files, and associated pulse files, is planned. Repeatability of *RA*-correlated polarized pulse pair anomalies seems important.

## VII. ACKNOWLEDGMENTS

Special thanks are given to Deep Space Exploration Society members, the workers of the SETI Institute, Breakthrough Listen and the U.C. Berkeley SETI Research Center, for work, ideas and helpful guidance. Special thanks are given to the workers of the Green Bank Observatory. Special thanks are given to family and friends for helpful advice.

## VIII. REFERENCES

[1] W. J. Crilly Jr, *An interstellar communication method: system design and observations*, arXiv: 2105.03727v1, May 8, 2021

[2] C. E. Shannon, W. Weaver, *The Mathematical Theory of Communication*, Urbana and Chicago, IL: University of Illinois Press, pp. 97–103, 1949

[3] D. G. Messerschmitt, *End-to-end interstellar communication system design for power efficiency*, arXiv 1305.4684v2, pp. 26–27, 70–74, 86–106, 201–223, 2018

[4] J. G. Proakis, *Digital Communications*, 4th Ed., New York, NY: McGraw-Hill, pp. 800–809, 2001

[5] A. Gelman, J. B. Carlin, H. S. Stern, D. B. Dunson, A. Vehtari, D. B. Rubin, *Bayesian Data Analysis*, 3rd Ed., Boca Raton, FL: CRC Press, p. 197, 2014